\pdfoutput=1
\newif\ifFull
\Fulltrue
\ifFull
\documentclass[11pt]{article}
\else
\documentclass{llncs}
\fi

\usepackage{graphicx}
\usepackage{amsmath}
\ifFull
\usepackage{amsthm}
\usepackage{fullpage}
\usepackage{hyperref}
\fi
\usepackage{amssymb}
\usepackage{url}
\ifFull
\fi

\ifFull
\newtheorem{theorem}{Theorem}
\newtheorem{lemma}[theorem]{Lemma}

\newtheorem{definition}[theorem]{Definition}
\fi

\newcommand{\R}{{\bf R}}
\newcommand{\Z}{{\bf Z}}
\newcommand{\cycle}{\mathrm{cycle}}
\newcommand{\level}{\mathrm{level}}
\newcommand{\parent}{\mathrm{parent}}
\newcommand{\child}{\mathrm{child}}
\newcommand{\depth}{\mathrm{depth}}

\newcommand{\head}{\mathrm{head}}
\newcommand{\tail}{\mathrm{tail}}
\newcommand{\relativeDepth}{\mathrm{relativeDepth}}
\newcommand{\boundary}{\mathrm{boundary}}
\newcommand{\radialOrder}{\mathrm{radialOrder}}

\newcommand{\superlevel}{\mathrm{superlevel}}

\ifFull
\newcommand{\leaveout}[1]{#1}  % full version
\newcommand{\Leaveout}[2]{#1}  % full version
\else
\newcommand{\leaveout}[1]{ }
\newcommand{\Leaveout}[2]{#2}
\let\oldendproof\endproof
\def\endproof{\qed\oldendproof}
\renewcommand{\subsection}[1]{\paragraph{#1.}}
\fi

\title{Succinct Greedy Geometric Routing \\
in the Euclidean Plane\thanks{This work was supported by
NSF grants 0724806, 0713046, 0830403, and ONR grant
N00014-08-1-1015.}\\[-12pt]}

\ifFull
\author{Michael T. Goodrich \\
Dept. of Computer Science \\
Univ. of California, Irvine \\
\url{http://www.ics.uci.edu/~goodrich/}
\and 
Darren Strash \\
Dept. of Computer Science \\
Univ. of California, Irvine \\
\url{http://www.ics.uci.edu/~dstrash/}
}
\date{}
\else
\author{Michael T. Goodrich \and Darren Strash}
\institute{Computer Science Department, University of California, Irvine, USA.}
\fi

\begin{document}

\maketitle

\ifFull
\pagestyle{plain}
\setcounter{page}{1}
\fi

\begin{abstract}
\ifFull
In greedy geometric routing, messages are passed in a network
embedded in a metric space according to the greedy strategy of
always forwarding messages to nodes that are closer to the destination.
\fi
We show that greedy geometric routing schemes exist for the Euclidean
metric in $\R^2$, for 3-connected planar graphs, 
with coordinates that can be represented \emph{succinctly}, that
is, with $O(\log{n})$ bits, where $n$ is the number of vertices
in the graph.  
\ifFull Moreover, our embedding strategy
introduces a coordinate system for $\R^2$ that supports distance
comparisons using our succinct coordinates. \fi
\ifFull
Thus, our scheme can
be used to significantly reduce bandwidth, space, and header size
over other recently discovered greedy geometric routing implementations
for $\R^2$.
\fi
\end{abstract}

\section{Introduction}

\ifFull
In an intriguing confluence of computational geometry and networking, 
\emph{geometric routing} has shown how simple
geometric rules can replace cumbersome routing tables
to facilitate effective message passing in a network
(e.g., see~\cite{bmsu-rgdah-01, eg-sggdh-08,kk-gpsr-00,%
k-gruhs-07,kwzz-gacr-03,kwz-aogma-02,kwz-wcoac-03}).
\fi
Geometric routing algorithms
perform message passing using geometric information
stored at the nodes and edges of a network.
\ifFull
For example,
geometric information could come from the latitude and longitude GPS 
coordinates of the nodes in a wireless sensor network or 
this information could come from an embedded doubly-connected 
edges list representation of
a planar subgraph of such a network.
Indeed, in one of the early works on the subject, 
Bose \textit{et al.}~\cite{bmsu-rgdah-01}
show how to do geometric routing in an embedded planar subgraph of a wireless
sensor network by using a geometric subdivision traversal 
algorithm of Kranakis {\it et al.}~\cite{kranakiscompass}, which was first
introduced in the computational geometry literature.
\fi

\subsection{Greedy Geometric Routing}
Perhaps the simplest routing rule is the 
\emph{greedy} one:
\begin{itemize}
\item
If a node $v$ receives a message $M$ 
intended for a destination $w\not= v$, 
then $v$ should forward $M$ to a neighbor that is
closer to $w$ than $v$ is.
\end{itemize}
This rule can be applied in any metric space, of course, but simple
and natural metric spaces are preferred over cumbersome or artificial
ones.
\ifFull

The greedy routing rule traces its roots back to the original
``degrees-of-separation'' small-world experiment of 
Milgram~\cite{m-swp-67}, where he asked randomly chosen individuals 
to forward 296 letters, initiating in Omaha, Nebraska and Wichita, 
Kansas, all intended for a lawyer in Boston, using the rule that 
requires each letter to be forwarded to an acquaintance that is 
closer to the destination.

In the modern context, researchers are interested in solutions that
use a paradigm introduced by Rao {\it et al.}~\cite{rrpss-grli-03} of
doing greedy geometric routing in geometric graphs that
assigns virtual coordinates in a metric space
to each node in the network, rather
than relying on physical coordinates.
For example, GPS coordinates 
may be unavailable for some sensors or the physical coordinates of
network nodes may be known only to a limited degree of certainty.
\fi
Thus, we are interested in greedy routing schemes that assign 
network nodes to virtual coordinates in a natural metric space.

\ifFull
Interestingly, the feasibility of the greedy routing
rule depends heavily on the geometry of the underlying metric space
used to define the notion of ``closer to the destination.''
For example, it is easy to see that star graphs (consisting
of a central vertex adjacent to every node in an arbitrarily large 
independent set)
cannot support greedy geometric routing in any 
fixed-dimensional Euclidean space. 
By a simple packing argument, there has to be two members of the
large independent set, in such a graph,
that will be closer to each other than the central
vertex.
Likewise, even for bi-connected or tri-connected planar graphs
embedded in $\R^2$, a
network may have ``holes'' where greedy routing
algorithms could get ``stuck'' in a local metric minimum
(e.g., see Funke~\cite{f-thdws-05} for
related work on hole detection in sensor networks).
Alternatively, several researchers (e.g.,
see~\cite{eg-sggdh-08,k-gruhs-07,m-adgra-07}) have shown that greedy
geometric routing is possible, for any connected graph,
in fixed-dimensional hyperbolic spaces.
Our interest in this paper, however, is on 
greedy geometric routing in $\R^2$ under the Euclidean metric,
since this space more closely matches the geometry of 
wireless sensor networks.
\fi

Interest in greedy geometric routing in fixed-dimensional Euclidean
spaces has expanded greatly since the work by Papadimitriou and
Ratajczak~\cite{pr-ocrgr-05}, who showed that any 3-connected planar graph
can be embedded in $\R^3$ so as to support greedy geometric routing. 
Indeed, their conjecture that such embeddings are possible in $\R^2$
spawned a host of additional papers (e.g., \ifFull see~\cite{afg-acgdt-08,%
cgw-dcvc-07,d-gdt-08,eg-sggdh-08,lp-oelia-08,m-adgra-07}).
\else
see~\cite{afg-acgdt-08,d-gdt-08,eg-sggdh-08,lp-oelia-08,m-adgra-07}).
\fi
Leighton and Moitra~\cite{lm-srgem-08} settled this conjecture by 
giving an algorithm to produce a greedy embedding of any 
3-connected planar graph in $\R^2$, and a similar result 
was independently found by Angelini {\it et al.}~\cite{afg-acgdt-08}. 
Greedy embeddings in $\R^2$ were previously known 
\ifFull only \fi for \ifFull graphs containing power 
diagrams~\cite{cgw-dcvc-07}, \fi graphs containing Delaunay 
triangulations~\cite{lp-oelia-08}, and existentially 
(but not algorithmically) for triangulations~\cite{d-gdt-08}.

\subsection{Succinct Geometric Routing}
In spite of their theoretical elegance, these results settling the
Papadimitriou-Ratajczak conjecture have an unfortunate drawback, in
that the virtual coordinates of nodes in these solutions require
$\Omega(n\log n)$ bits each in the worst case.
These space inefficiencies reduce the applicability of
these results for greedy geometric routing, since one could 
alternatively keep routing tables of size $O(n\log n)$ bits 
at each network node to support message passing.
Indeed, such routing tables would allow for network nodes to be identified
using labels of only $O(\log n)$ bits each, which would significantly
cut down on the space, bandwidth, and packet header size 
needed to communicate the
destination for each packet being routed.
Thus, for a solution to be effectively solving the routing problem
using a greedy geometric routing  scheme, we desire 
that it be
\emph{succinct}, that is, it should use
$O(\log{n})$ bits per virtual coordinate.
Succinct greedy geometric routing schemes are known for
fixed-dimensional hyperbolic spaces~\cite{eg-sggdh-08,m-adgra-07}, 
but we are unaware of any prior work on succinct greedy 
geometric routing in fixed-dimensional Euclidean spaces.
\ifFull
We are therefore interested in this paper in a method for 
succinct greedy geometric routing in $\R^2$, with distance comparisons being
consistent with the standard Euclidean $L_2$ metric.

\subsection{Additional Related Prior Work}
In addition to the greedy geometric routing schemes referenced above,
there is a hybrid scheme, for example, as outlined by Karp and
Kung~\cite{kk-gpsr-00}, which
combines a greedy routing strategy with face
routing~\cite{bmsu-rgdah-01}.
Similar hybrid schemes were subsequently studied by several other
researchers 
(e.g., see~\cite{fs-odgfc-06,kwzz-gacr-03,kwz-aogma-02,kwz-wcoac-03}).
An alternative hybrid augmented greedy
scheme is introduced by Carlsson and Eager~\cite{ce-negrw-07}.
In addition, Gao {\it et al.}~\cite{gghzz-gsrmn-05} show how to
maintain a geometric spanner in a mobile network so as to support
hybrid routing schemes.
Although such schemes are \emph{local}, in that routing decisions can
be made at a node $v$ simply using information about $v$'s neighbors,
we are interested in this paper in routing methods that are purely
greedy.

As mentioned above,
Rao {\it et al.}~\cite{rrpss-grli-03} introduce the idea of doing
greedy geometric routing 
using virtual coordinates, although they make no theoretical guarantees,
and Papadimitriou and Ratajczak~\cite{pr-ocrgr-05}
are the first to prove such a method exists in $\R^3$,
albeit with a non-standard metric.
In addition,
we also mentioned above how
Leighton and Moitra~\cite{lm-srgem-08} 
and Angelini {\it et al.}~\cite{afg-acgdt-08}
have settled the Papadimitriou-Ratajczak conjecture, albeit with 
solutions that are not succinct.
Moreover,
the only known succinct greedy geometric routing schemes 
are for fixed-dimensional hyperbolic spaces~\cite{eg-sggdh-08,m-adgra-07}.
Thus, there does not appear to be any prior work on 
succinct greedy geometric routing in $\R^2$ using the standard
Euclidean $L_2$ metric.

The problem of constructing succinct greedy
geometric routing schemes in $\R^2$ is related to the general
area of compressing geometric and topological data for networking
purposes. Examples of such work includes the compression
schemes of Suri {\it et al.}~\cite{ssw-ctdrt-03} 
for two-dimensional routing tables,
and the coordinate and mesh compression work of 
Isenburg {\it et al.}~\cite{ils-lcpfp-05}.
We should stress, therefore, that we are not primarily interested
in this paper in compression schemes for greedy geometric routing; we
are interested primarily in coordinate systems for greedy routing, since they
have a better applicability in distributed settings.
In particular, we are not interested in a compression scheme where
the computation of the coordinates in $\R^2$ of a network node $v$
depends on anything other than a succinct label for $v$.
That is,
we want a succinct \emph{coordinate system}, not simply an efficient
compression scheme that supports greedy routing.
Indeed, we show that succinct compression schemes are trivial, given
known Euclidean greedy geometric routing
methods~\cite{afg-acgdt-08,lm-srgem-08}.

Another area of related work is on methods for routing in geometric
graphs, such as road networks
(e.g., see~\cite{bfss-frrn-07,gh-cspasm-05,hsww-csuts-05,%
kp-sppls-06,ss-hhhes-05,sv-speg-86,zn-spaeu-98}).
For example, 
Sedgewick and Vitter~\cite{sv-speg-86} 
and Goldberg and Harrelson~\cite{gh-cspasm-05}
study methods
based on applying AI search algorithms, and
Bast {\it et al.}~\cite{bfss-frrn-07} explore routing methods based on the
use of transit nodes.
In this related work, the coordinates of the network nodes are fixed 
geometric points, whereas, in the greedy geometric routing problems we
study in this paper, vertices are assigned virtual coordinates so
as to support greedy routing.
\fi

\subsection{Our Results}
We provide a succinct greedy geometric routing scheme for 3-connected
planar graphs in $\R^2$. At the heart of our scheme is a new greedy 
embedding for 3-connected planar graphs in $\R^2$ which exploits the 
tree-like topology of a spanning (Christmas cactus) subgraph. Our
embedding allows us to form a coordinate system which uses $O(\log{n})$ bits 
per vertex, and allows distance comparisons to be done just using our 
coordinate representations consistently with the Euclidean metric.
\ifFull
Although we are primarily interested in such a coordinate system for greedy
geometric routing, we also give a simple global 
compression scheme for greedy geometric routing, based on the
approach of Leighton and Moitra~\cite{lm-srgem-08} 
and Angelini {\it et al.}~\cite{afg-acgdt-08}, which achieves $O(\log n)$
bits per vertex, which is asymptotically optimal.

\fi
\ifFull
Our coordinate scheme for greedy geometric routing in a graph $G$ 
is based on a three-phase approach. In the first phase, we find a 
spanning subgraph, $C$, of $G$, called a Christmas cactus 
graph~\cite{lm-srgem-08}. In the second phase, we find a 
graph-theoretic dual to $C$, which is a tree, $T$, and we form a 
heavy path decomposition on $T$.
Finally, in the third phase, we show how to use $T$ and $C$ to embed
$G$ in $\R^2$ to support greedy routing with coordinates that can be
represented using $O(\log^2{n})$ bits, and then we show how this can
be further reduced to $O(\log n)$ bits per node.
\fi

\section{Finite-Length Coordinate Systems}
Let us begin by formally defining what we mean by a coordinate
system, and how that differs, for instance, from a simple compression scheme.
Let $\Sigma$ be an alphabet, and let $\Sigma^*$ a set of finite-length strings 
over $\Sigma$. 
We define a \emph{coordinate system} $f$ for a space $S$:

\begin{enumerate}
\item $f$ is a map, $f:\Sigma^* \rightarrow S$, which assigns
character strings to points of $S$.

\item $f$ may be \emph{parameterized}: 
the assignment of strings to 
points may depend on a fixed set of parameters.

\item $f$ is \emph{oblivious}:
the value of $f$ on any given $x\in \Sigma^*$ must depend
only on $f$'s parameters and $x$ itself. It cannot
rely on any other character 
strings in $\Sigma^*$, points in $S$, or other values of $f$. 
\end{enumerate}

\ifFull
Clearly, this is a computationally-motivated definition of 
a coordinate system, since real-world computations performed on 
actual points must use finite representations of those points.
This is an issue and theme
present, for instance,
in computational geometry (e.g.,
see~\cite{%
bcddprty-pcet-99,by-ag-98,%
bepp-cegpu-97,by-eeesd-00,bkmnsu-egcl-95,%
em-sstcd-90,fv-eeacg-93,gght-srlse-97,gm-rad-98,ils-lcpfp-05,ph-srr-01,%
ps-cgi-90,s-apfpa-97,ssw-ctdrt-03}).
Note also that
our definition can be used to define finite versions of 
all the usual coordinate systems, since it allows for the use of symbols
like ``$\pi$'', ``$/$,'' and $k$-th root symbols. 
Thus, it supports finite coordinates using rational and algebraic
numbers, for example.
In addition, note that it supports points in non-Cartesian coordinate
systems, such as a finite-length polar coordinate system,
in that we can allow 
strings of the form ``$(x,y)$'' where 
$x$ is a string representing a value $r\in \R^{+}$ and $y$ 
is a string representing a value $\theta\in [0,2\pi)$, which may even use
``$\pi$''. It also allows for non-unique representations, like the 
homogeneous coordinate system for $\R^2$, which uses triples
of strings, with each triple representing a point in the Euclidean 
plane, albeit in a non-unique way.
\fi
If $f$ is lacking property 3, we prefer to think of $f$ 
as a \emph{compression scheme}.
\ifFull 
Examples of compression schemes are 
mappings that use look-up tables, which are built incrementally based on 
sequences of previous point assignments~\cite{ils-lcpfp-05}.
Given a compression scheme $f:\Sigma_f^*\rightarrow S$, 
note that it is possible to
construct a coordinate system $f':\Sigma_{f'}^*\rightarrow S$ by 
augmenting strings in $\Sigma_f^*$ with the data required 
to evaluate $f$ (such as the assignments of other points in a set of
interest).
\fi

\section{Greedy Routing in Christmas Cactus Graphs}
\label{section:greedy-routing}
Our method is a non-trivial adaptation of
the Leighton and Moitra scheme~\cite{lm-srgem-08}, so we begin
by reviewing some of the ideas from their work.

A graph $G$ is said to be a \emph{Christmas cactus graph} if: (1) each
edge of $G$ is in at most one cycle, (2) $G$ is connected, and (3) removing 
any vertex disconnects $G$ into at most two components.
%
%A \emph{cactus graph} is a graph where each edge is contained in at most one 
%cycle and a \emph{Christmas cactus graph} 
%is a connected cactus graph such that the removal of any vertex 
%disconnects the graph into at most 2 connected components. 
For ease of discussion, we consider any edge in a Christmas cactus graph that 
is not in a simple cycle to be a simple cycle itself (a 2-cycle); hence, 
every edge in is in exactly one simple cycle.
The \emph{dual tree} of a Christmas cactus graph $G$ is a tree 
containing a vertex for each simple cycle in $G$ with an edge 
between two vertices if their corresponding cycles in $G$ share a vertex. 
Rooting the dual tree at an arbitrary vertex creates what we call a \emph{depth tree}.\ifFull
(See Fig.~\ref{fig:cactus}.)

\begin{figure}[!hbt]
%\vspace*{-12pt}
\begin{center}
\begin{tabular}{cc}
\includegraphics[scale=0.7]{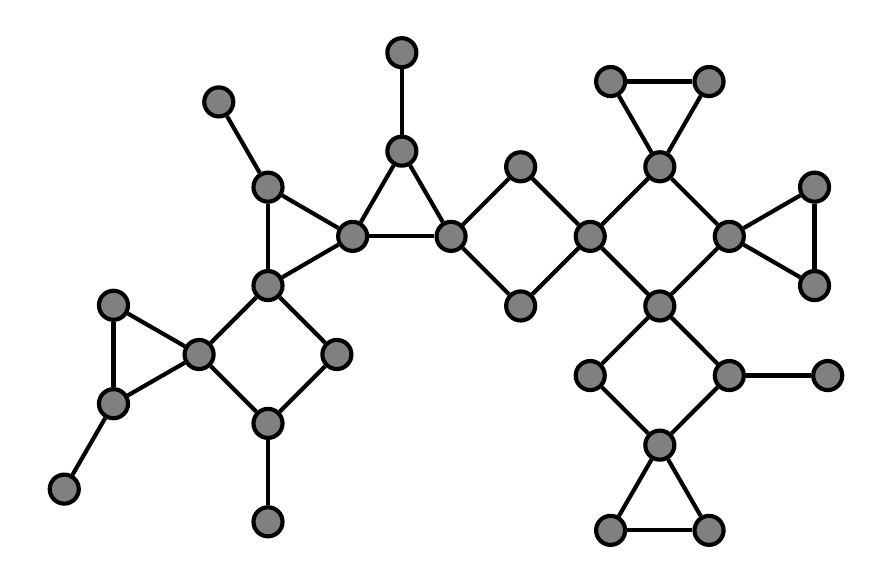} &
\includegraphics[scale=0.7]{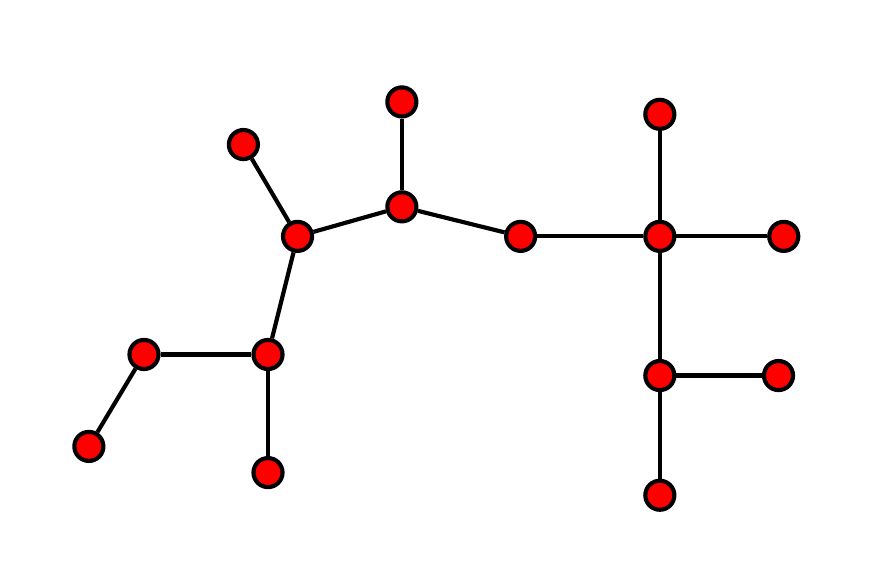} \\
(a) & (b)
\end{tabular}
\end{center}
%\vspace*{-12pt}
\caption{(a) A Christmas cactus graph and (b) its dual tree.}
\label{fig:cactus}
\end{figure}

\fi

Having a depth tree allows us to apply the rooted tree terminology to
cycles in $G$. In particular: \emph{root}, \emph{depth}, \emph{parent}, \emph{child}, 
\emph{ancestor}, and \emph{descendant} all retain their familiar definitions. 
We define the \emph{depth} of a node $v$ to be the minimum depth of any 
cycle containing $v$. The unique node that a cycle $C$ shares with its parent 
is called the \emph{primary node} of $C$. Node $v$ is a \emph{descendant} of 
a cycle $C$ if $v$ is in a cycle that is a descendant of $C$ and $v$ is not 
the primary node of $C$. Node $v$ is a \emph{descendant} of node $u$ if removing
neighbors of $u$ with depth less than or equal to $u$ leaves $u$ and $v$ in
the same component.

\subsection{Greedy Routing with a Christmas Cactus Graph Embedding}
\ifFull
Leighton and Moitra~\cite{lm-srgem-08} show that every 3-connected planar 
graph contains a spanning Christmas cactus subgraph and that every Christmas 
cactus graph has a greedy embedding in $\R^2$, which together imply
that 3-connected planar graphs have greedy embeddings in $\R^2$.
\fi
Working level by level in a depth tree, Leighton and Moitra~\cite{lm-srgem-08} 
embed the cycles of a Christmas cactus graph on semi-circles of increasing 
radii, centered at the origin. Within the embedding we say that vertex 
$u$ is above vertex $v$ if $u$ is embedded farther from the origin than 
$v$, and we say that $u$ is to the left of $v$ if $u$ is embedded in the positive
angular direction relative to $v$. We can define below and right similarly. 
These comparisons naturally give rise to
directions of movement between adjacent vertices in the embedding: 
up, down, left, and right.
\begin{figure}[!b]
%\vspace*{-24pt}
\begin{center}
\begin{tabular}{c@{\hspace*{5em}}c}
    \includegraphics[scale=0.7,viewport=0cm 3.5cm 7cm 8.5cm, clip=true]{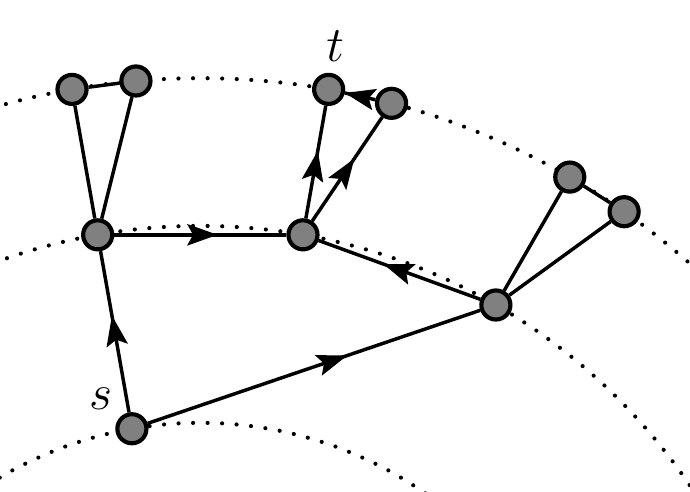}
    &
    \includegraphics[scale=0.7,viewport=0cm 3.5cm 7cm 8.5cm, clip=true]{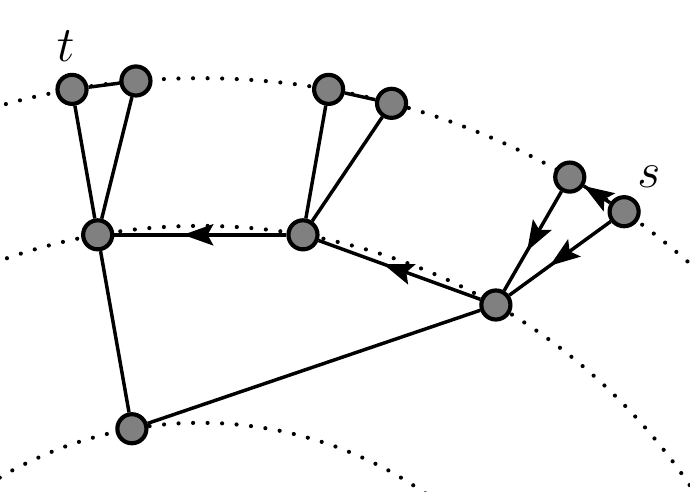} \\%[-8pt]
    (a) & (b)
    \end{tabular}
    \end{center}
%    \vspace*{-12pt}
    \caption{ Arrows indicate valid greedy hops. 
    (a) Descendants of $s$ can be reached by a simple path of 
        up and right hops, up and left hops, or
        a combination of the two.
    (b) If $t$ is not a descendant of $s$, then we route down and (left or right)
        in the direction of $t$ until we reach an ancestor of $t$.
    }
    \label{fig:subcactus_at_s}
    \label{fig:cactus_lca}
%    \vspace{-12pt}
\end{figure}

Routing from start vertex $s$ to a terminal vertex $t$ in a Christmas
cactus graph embedding can be broken down into two cases: (1) $t$ is 
a descendant of $s$, and (2) $t$ is not a descendant of $s$.
\begin{enumerate}
    \item As shown in Fig.~\ref{fig:subcactus_at_s}(a), if $t$ is a descendant 
          of $s$, then we can route to $t$ by a simple path of up and  
          right hops, up and left hops, or a 
          combination of the two. 

    \item As shown in Fig.~\ref{fig:cactus_lca}(b), if $t$ is not a descendant
          of $s$, then we route to the least common (cycle) ancestor of 
          $s$ and $t$. Suppose, without loss of generality, that $t$ is to 
          the left of $s$, then we can reach this cycle by a sequence of 
          down and left hops. Once on the cycle, we can 
          move left until we reach an ancestor of $t$. Now we are back in 
          case 1. 
\end{enumerate}
\ifFull
\subsection{A Succinct Compression Scheme}
Using the Christmas cactus graph embedding discussed above, we can assign 
succinct integer values to each vertex, allowing us perform greedy 
routing according to the Euclidean $L_2$ metric. Our embedding 
$f:V(G)\rightarrow \Z_{n}^3$ produces a triple of the following integers: 
$\radialOrder(v)$: the number of vertices to the right of $v$;
$\level(v)$: the number of semi-circles between the vertex 
and the origin, excluding the semi-circle that $v$ is embedded on; and
$\boundary(v)$: the smallest $\radialOrder$ value of all vertices that are 
descendants of $v$. The Leighton-Moitra embedding has the 
property that all descendants of $v$ fall between $v$ and the 
vertex embedded immediately to the right of $v$ on the same level as $v$. 
Since each element of the triple can take on values in the range $[0,n]$,
the triple can be stored using $O(\log n)$ bits.

We can implement each step of the routing scheme using only the 
triples of $s$, the neighbors of $s$, and $t$. Queries of the form 
\emph{$u$ is left/right of $v$}, involve a straightforward 
comparison of the $\radialOrder$ element of the triple. Likewise for \emph{$u$ is 
above/below $v$}, using $\level$. The same comparisons can be used to determine 
which neighbors of $s$ are a left, right, down, or 
up move away. Finally, queries of the form \emph{$u$ is a descendant
of $v$} are true if and only if $\boundary(v) \leq \radialOrder(u) \leq 
\radialOrder(v)$ and $\level(v) \leq \level(u)$. 

To extend this routing scheme to graphs that have a spanning Christmas cactus 
subgraph, we need to ensure that the routing scheme
does not fail by following edges that are not in the Christmas cactus subgraph. 
Since the Christmas cactus graph has bounded degree $4$, for a node $v$, we can 
store the triples of neighbors of $v$ in the Christmas cactus graph, in addition
to storing the triple for $v$, and only allow our greedy routing scheme to 
choose vertices that are neighbors in the Christmas cactus subgraph. Storing
these extra triples in the coordinate does not increase its asymptotic 
bit-complexity.

This routing scheme is greedy according to the Euclidean coordinates of the 
vertices, using the Euclidean $L_2$ metric. Unfortunately, if we only 
have access to the integer triples then it is not obvious that there 
is any metric that we can define that will satisfy the definition 
for greedy routing using just these integer values. Therefore, we must 
concede that, while this routing scheme fulfills the spirit of greedy 
routing, it is not greedy routing in the strictest sense. This is an 
example of a compression scheme and not a coordinate system.
\else
This routing scheme immediately gives rise to a simple 
succinct compression scheme for 3-connected planar graphs, which we discuss in
the full version of this paper. 
\fi

\section{Toward a Succinct Greedy Embedding} 
Given a 3-connected planar graph, we can find a spanning Christmas cactus
subgraph in polynomial time~\cite{lm-srgem-08}. Therefore, we restrict our
attention to Christmas cactus graphs. Our results apply to 
3-connected planar graphs with little or no modification. In this section, we construct a novel 
greedy embedding scheme for any Christmas cactus graph in $\R^2$. We 
then build a coordinate system from
our embedding and show that the coordinates can be represented using 
$O(\log^2{n})$ bits. In the next section, we show how to achieve an
optimal $O(\log{n})$-bit representation.

%\leaveout{
%\begin{figure}[!ht]
%\begin{center}
%\includegraphics[width=2.0in]{embed_overview.png}
%\end{center}
%\caption{A view of our embedding between super levels. Blue triangles represent 
%cycles on the same heavy path, other triangles begin a new heavy path.}
%\end{figure}
%}

\subsection{Heavy Path Decompositions}

We begin by applying the Sleator and Tarjan~\cite{SleTar-JCSS-83} 
\emph{heavy path decomposition} to the depth tree $T$ for $G$.
\begin{definition} Let $T$ be a rooted tree. For each node $v$ in $T$, 
let $n_{T}(v)$ denote the number of descendants of $v$ in $T$, including $v$. 
For each edge $e = (v,\parent(v))$ in $T$, label $e$ as a heavy edge 
if $n_{T}(v) > n_{T}(\parent(v))/2$. Otherwise, label $e$ as a light edge. 
Connected components of heavy edges form paths, called heavy paths. 
Vertices that are incident only to light edges are considered to be 
zero-length heavy paths. We call this the 
{\bfseries heavy path decomposition} of $T$.
\end{definition}
For ease of discussion, we again apply the terminology from nodes in $T$ to
cycles in $G$. A cycle in $G$ is on a heavy path $H$ if its dual node in $T$
is on $H$. Let $H$ be a heavy path in $T$. We say that $\head(H)$ is the 
cycle in $H$ that has minimum depth, we define $\tail(H)$ similarly. 
Let $C_1$ and $C_2$ be two cycles such that $C_1 = \parent(C_2)$ and let 
$\{p\} = V(C_1) \cap V(C_2)$.
If $C_1$ and $C_2$ are on the same heavy path then we call $p$ a \emph{turnpike}.
If $C_1$ and $C_2$ are on different heavy paths (where $C_1=\tail(H_1)$ and 
$C_2 = \head(H_2)$) then we call $p$ an \emph{off-ramp} for $H_1$ and the vertices 
$v\in V(C_2) \setminus \{p\}$ \emph{on-ramps} for $H_{2}$. 

\subsection{An Overview of Our Embedding Strategy}
Like Leighton and Moitra~\cite{lm-srgem-08}, we lay the cycles 
from our Christmas cactus graph on concentric semi-circles of radius 
$1=R_0 < R_1 < R_2 \ldots$; however, our embedding has the 
following distinct differences: we have $\Theta(n\log n)$ semi-circles
instead of $O(n)$ semi-circles, on-ramps to heavy paths are embedded on special 
semi-circles which we call \emph{super levels},
turnpikes are placed in a predefined position when cycles are embedded, 
and the radii of semi-circles can be computed without knowing the topology 
of the particular Christmas cactus graph being embedded. 
\ifFull
Since the path from the root to any leaf in the depth tree 
contains $O(\log{n})$ heavy paths, our embedding has $O(\log{n})$ 
of super levels. Between super levels we lay out the non-trivial 
heavy paths on \emph{baby levels}. \fi

To make our embedding scheme amenable to a proof by induction, we
modify the input Christmas cactus graph. After constructing a greedy embedding of 
this modified graph, we use it to prove that we have a greedy 
embedding for the original graph.

\subsection{Modifying the Input Christmas Cactus Graph}
Given a Christmas cactus graph $G$ on $n$ vertices, we choose a depth 
tree $T$ of $G$, and compute the heavy path decomposition of $T$. 
For a cycle $C$ on a heavy path $H$, we define $\relativeDepth(C)$ to be 
$\depth(C) - \depth(\head(H))$. For each $C_1$, $C_2 = \child(C_1)$ forming 
a light edge in $T$, let $\{p\} = V(C_1) \cap V(C_2)$. Split $p$ into two 
vertices $p_1$ and $p_2$ each on their own cycle, and connect $p_1$ to $p_2$ 
with a path of $n-1-\relativeDepth(C_1)$ edges. The new graph $G'$ 
is also a Christmas cactus graph, and our new depth tree $T'$ looks 
like $T$ stretched out so that heads of heavy paths (from $T$) are 
at depths that are multiples of $n$. (See Fig.~\ref{fig:heavy}.) 
\leaveout{
\begin{figure}[!t]
\begin{center}
\begin{tabular}[b]{cc}
    \includegraphics[scale=0.7]{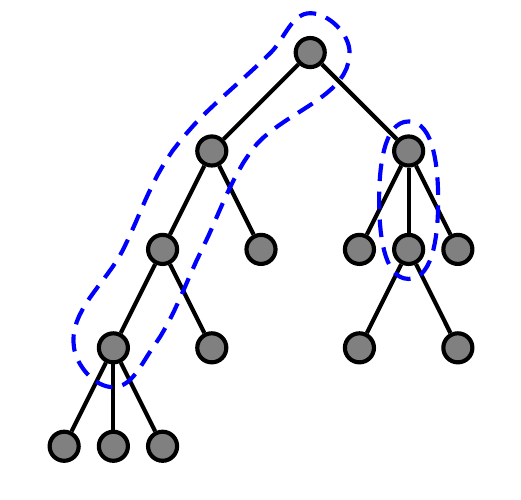} &
    \includegraphics[scale=0.7,clip=true,viewport=6cm 1.5cm 12.5cm 11cm]{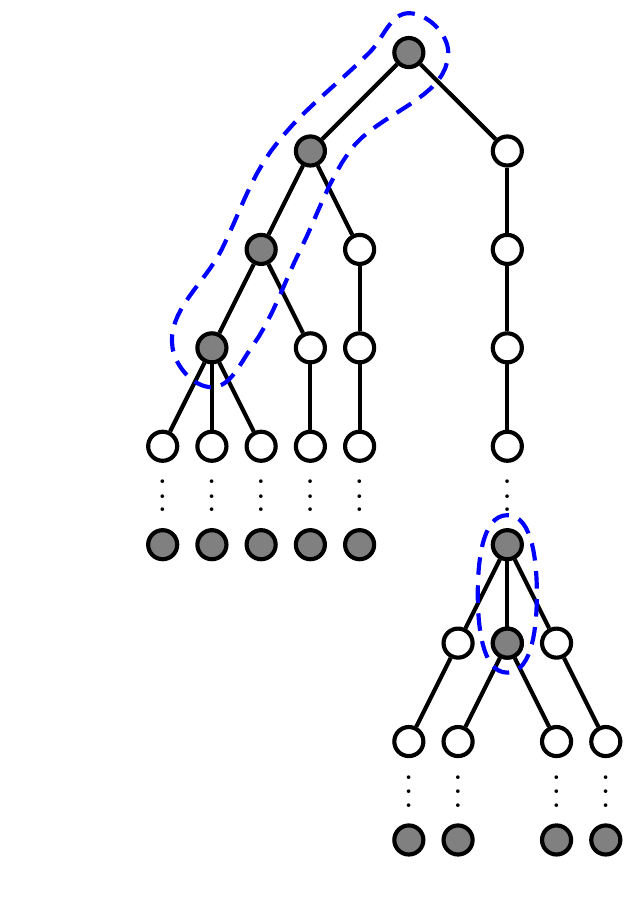} \\
    (a) & (b)
\end{tabular}
    \caption{(a) A depth tree $T$ with positive-length heavy paths highlighted, and 
             (b) the new depth tree $T'$ after the modification procedure.}
    \label{fig:heavy}
\end{center}
\end{figure}
}
We continue to call the paths copied from $T$ heavy paths (though they do 
not form a heavy path decomposition of $T'$), and the newly inserted edges 
are \emph{dummy} edges.

\subsection{Embedding the Modified Christmas Cactus Graph in $\R^2$}
Given a Christmas cactus graph $G$ on $n$ vertices, run the modification 
procedure described above and get $G'$ and $T'$. We embed $G'$ in 
phases, and prove by induction that at the end of each phase
we have a greedy embedding of an induced subgraph of $G'$.

\begin{lemma}[Leighton and Moitra~\cite{lm-srgem-08}]
\label{lemma-geom}
\ifFull
If the coordinates 
\begin{align*}
c &= (0,1 + z) \\
b &= (-\sin \beta, \cos \beta) \\
a &= (-(1 + \epsilon)\sin(\beta - \alpha),(1+\epsilon)\cos(\beta - \alpha))
\end{align*}
are subject to the constraints
\begin{align*}
0 < &\alpha,\beta \leq \pi/2 \\
0 < &\epsilon \leq \frac{1 - \cos \beta}{6} \\
0 \leq &z \leq \epsilon \\
\sin \alpha \leq &\frac{\epsilon(1 - \cos \beta)}{2(1+\epsilon)}
\end{align*}
then $d(a,c) - d(b,c) \geq \epsilon^2 > 0$.
\else %
If points
$c = (0,1 + z)$, $b = (-\sin \beta, \cos \beta)$, and 
$a = (-(1 + \epsilon)\sin(\beta - \alpha),(1+\epsilon)\cos(\beta - \alpha))$
are subject to the constraints
$0 < \alpha \leq \pi/2$, $0 < \beta \leq \pi/2$, 
$0 < \epsilon \leq (1 - \cos \beta)/6$,
$0 \leq z \leq \epsilon$, and
$\sin \alpha \leq \frac{\epsilon(1 - \cos \beta)}{2(1+\epsilon)}$ %(\epsilon(1 - \cos \beta))/(2(1+\epsilon))$
then $d(a,c) - d(b,c) \geq \epsilon^2 > 0$.
\fi
\end{lemma}

We begin by embedding the root cycle, $C = (v_0,\ldots,v_{k-1})$, of $T'$. 
We trace out a semi-circle of radius $R_0=1$ centered at the origin and 
divide the perimeter of this semi-circle into $2n+1$ equal arcs. We allow
vertices to be placed at the leftmost point of each arc, numbering these 
positions $0$ to $2n$. We place vertices $v_0,\ldots,v_{k-1}$ clockwise into 
any $k$ distinct positions, reserving position $n$ for $C$'s turnpike. If $C$ 
does not have a turnpike, as is the case if $C$ is a dummy edge or the tail 
of a heavy path, then position $n$ remains empty. The embedding of $C$ is 
greedy\Leaveout{}{\ (proof omitted here)}.

\leaveout{
%\begin{figure}[!htb]
%\centering
%\includegraphics[width=2.0in]{root_embed.png}
%\caption{The $2n+1$ positions where vertices from the root cycle can be 
%embedded.}
%\label{fig:root_embed}
%\end{figure}

\begin{proof} If $C$ is a 2-cycle, then 
the embedding of $C$ is greedy regardless of where the vertices are 
embedded. Otherwise, consider each segment $su\neq v_0v_{k-1}$. The 
perpendicular bisector to $su$ does not intersect any of our
embedded vertices. $u$ is the neighbor of $s$ that is closer to every vertex
on the $u$ side of the perpendicular bisector. Since all such segments have this
property, the embedding of $C$ is greedy.
\end{proof}
}

\emph{Inductive Step:}
Suppose we have a greedy embedding all cycles in $T'$ up to depth $i$,
call this induced subgraph $G_i'$. We show that the embedding can
be extended to a greedy embedding of $G_{i+1}'$. Our proof relies on 
two values derived from the embedding of $G_i'$. 

\begin{definition}
Let $s$, $t$ be any two distinct vertices in $G_i'$ and 
fix $n_{s,t}$ to be a neighbor of $s$ such that $d(s,t) > d(n_{s,t},t)$. 
We define $\delta(G_i')=\min_{s,t}\{d(s,t) - d(n_{s,t},t)\}$.
\end{definition}

We refer to the difference $d(s,t) - d(n_{s,t},t)$ as the \emph{delta value} 
for distance-decreasing paths from $s$ to $t$ through $n_{s,t}$.

\begin{definition}
Let $\beta(G_i')$ to be the minimum (non-zero) angle 
that any two vertices in the embedding of $G_i'$ form with the origin.
\end{definition}
\begin{figure}[!]
\vspace*{-28pt}
\begin{center}
\includegraphics[scale=0.7]{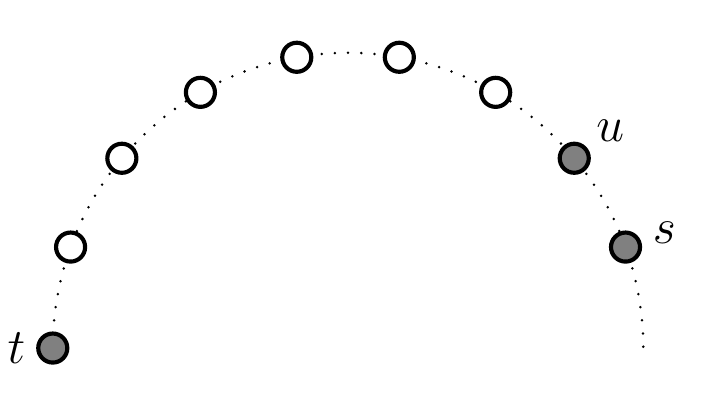}
\end{center}
\vspace*{-16pt}
\caption{$s$, $u$ and $t$ form a lower bound for $\delta(G_0')$.}
\vspace*{-16pt}
\label{fig:delta_0}
\end{figure}
Since we do not specify exact placement of all vertices, we cannot compute 
$\delta(G_0')$ and $\beta(G_0')$ exactly. We instead compute positive 
underestimates, $\delta_0$ and $\beta_0$, by considering hypothetical
vertex placements, and by invoking the following lemma.

\begin{lemma}
\label{lemma-delta}
Let $s$ and $u$ be two neighboring vertices embedded in the 
plane. If there exists a vertex $t$ that is simultaneously 
closest to the perpendicular bisector of $su$ (on the $u$ side), 
and farthest from the line $su$, then the delta value for $s$ to $t$
through $u$ is the smallest for any choice of $t$. 
\end{lemma}

Applying the above lemma to all hypothetical $s$, $u$, and $t$ placements for the
embedding of $G_0'$ leads to the underestimate 
$\delta_0 = 2 - \sqrt{2 + 2\cos{\frac{\pi}{2n+1}}} 
< d(s,t)-d(u,t) \leq \delta(G_0')$ where $s$, $u$, and $t$ are shown in Fig.~\ref{fig:delta_0}. Trivially, $\beta_0 = \frac{\pi}{2n+1} \leq \beta(G_0')$.

We now show how to obtain a greedy embedding of $G_{i+1}'$, given a greedy embedding of $G_i'$ and values $\delta_i$ and $\beta_i$.

Let $\epsilon_i = \min\{\delta_i/3, R_i\frac{1-\cos{\frac{2}{3}\beta_i}}{6}\}$.
Trace out a semi-circle of radius $R_{i+1} = R_i + \epsilon_i$ centered at 
the origin. Each cycle at depth $i+1$ of $T'$ has the form $C=(v, x_1,\ldots, x_m)$ 
where $v$, the primary node of $C$, has already been embedded on the $i$th 
semi-circle. We embed vertices $x_1$ to $x_m$ in two subphases: \\

\noindent\emph{Subphase 1} We first embed vertex $x_1$ from each $C$. 
Choose an orientation for $C$ so that $x_1$ is not a turnpike.\footnote{For 
the case where $C$ is a 2-cycle and $x_1$ is a turnpike we insert a
temporary placeholder vertex $p$ into $C$ with edges to $v$ and $x_1$, and treat
$p$ as the new $x_1$. We can later remove this placeholder by transitivity.} We place $x_1$ where the ray beginning at the origin 
and passing through $v$ meets semi-circle $i+1$. We now show that distance 
decreasing paths exist between all pairs of vertices embedded thus far. 

Distance decreasing paths between vertices in $G_i'$ are preserved by the 
induction hypothesis. For $t$ placed during this subphase: $t$ has a neighbor $v$ 
embedded on semi-circle $i$. If $s=v$ then $s$'s neighbor $t$ is strictly closer to $t$. 
Otherwise if $s\in G_i'$ then since $t$ is within distance $\delta_i/3$ of $v$, 
then $s$'s neighbor $u$ that is closer to $v$ is also closer to $t$. 
\ifFull
By definition of $\delta_i$, $d(s,v) \geq d(u,v) + \delta_i$.

Since $t$ is in the $\delta_i/3$-ball around $v$, 
$d(s,t) \geq d(s,v) - \delta_i/3$, and $d(u,t) \leq d(u,v) + \delta_i/3$.

Then,
\begin{align*}
    d(s,t) &\geq d(s,v) - \delta_i/3\\
           &\geq d(u,v) + \delta_i - \delta_i/3\\
           &\geq d(u,v) - \delta_i/3  + \delta_i - \delta_i/3\\
           &= d(u,t) + \delta_i/3\\
           &> d(u,t)
\end{align*}
Therefore, $s$'s neighbor $u$ that is closer to the primary node $v$ is also closer to $t$.
\fi
If $s$ was placed during this subphase then $s$ is within distance
$R_i\frac{1-\cos{\frac{2}{3}\beta_i}}{6}$ from its neighbor $v$, and
the perpendicular bisector of $sv$ contains $s$ on one side and 
every other vertex placed on the other side. Therefore $s$'s neighbor 
$v$ is closer to $t$.

The next subphase requires new underestimates, which we call $\delta^1_{i}$ 
and $\beta^1_i$. By construction, $\beta^1_i = \beta_i$. 
No $s$--$t$ paths within $G_i'$ decrease the delta value.
Paths from $s\in G_i'$ to $t$ placed in this subphase have delta value at 
least $\delta_i/3$ by design.
\ifFull This follows directly from the proof of greediness of this 
subphase.\fi
For paths from $s$ placed in this subphase, $s$'s neighbor $v$ is 
the closest vertex to the perpendicular bisector 
of $sv$ on the $v$ side. If we translate $v$ along the perpendicular bisector of $sv$ 
to a distance of $R_{i+1}$ from $sv$, this hypothetical point allows us to invoke 
Lemma~\ref{lemma-delta} to get an underestimate for the delta value of all 
paths beginning with $s$. Therefore, our new underestimate is: 
$\delta^1_{i} = \min\{\delta_i/3, \sqrt{R_{i+1}^2 + \epsilon_i^2} - R_{i+1}\}$.\\ 

\noindent\emph{Subphase 2} We now finish embedding each cycle 
$C = (v, x_1,\ldots x_m)$. Let the value $\alpha = \min\{\beta^1_{i}/3,$ $\delta^1_{i}/(3R_{i+1})\}$, 
s.t. $\sin{\alpha} \leq 
\frac{\epsilon_i(1-\cos{\frac{2}{3}\beta^1_i)}}{2(1+\epsilon_i)}$.  
Trace out an arc of length $R_{i+1}\alpha$ from the embedding of $x_1$, clockwise
along semi-circle $i+1$. We evenly divide this arc into $2n+1$ 
positions, numbered $0$ to $2n$. Position $0$ is already filled by $x_1$. 
We embed vertices in clockwise order around the arc in $m-1$ 
distinct positions; reserving position $n$ for $C$'s turnpike. If 
there is no such node, position $n$ remains empty.

This completes the embedding of $G_{i+1}'$. We show that the embedding of 
$G_{i+1}'$ is greedy. We only need to consider distance decreasing paths 
that involve a vertex placed during this subphase. For $t$ placed during
this subphase, $t$ is within distance $\delta^1_{i}/3$ from an $x_1$, 
therefore, all previously placed $s \neq x_1$ have a neighbor $u$ that is closer 
to $t$. If $s=x_1$ the $s$'s neighbor closer to $t$ is $x_2$. 
Finally, for $s$ placed during this subphase, let the cycle that 
$s$ is on be $C=(v, x_1,\ldots,x_m)$. For $s=x_i\neq x_m$,
since $\alpha \leq \beta^1_i/3$, the interior of the sector formed by 
$x_1$, $x_m$ and the origin is empty, therefore
$t$ is either on the $x_{i-1}$ side of the 
perpendicular bisector to $x_{i-1}x_i$ or on the $x_{i+1}$
side of the perpendicular bisector to $x_ix_{i+1}$.
If $s=x_m$ If $t$ is embedded to the left $s$, the closer neighbor 
is $x_{m-1}$. Otherwise, applying Lemma~\ref{lemma-geom},
our choice of $\sin{\alpha} \leq 
\frac{\epsilon_i(1-\cos{\frac{2}{3}\beta^1_i})}{2(1+\epsilon_i)}$
forces the perpendicular bisector to $sv$ to have $s$ on one side, and
all nodes to the right of $s$ on the other side. All cases are considered,
so the embedding of $G_{i+1}'$ is greedy. 

To complete the inductive proof, we must compute $\delta_{i+1}$ and $\beta_{i+1}$. 
Trivially, $\beta_{i+1} = \frac{\alpha}{2n} \leq \beta(G_{i+1}')$.
Distance decreasing paths between vertices placed before this subphase
will not update the delta value. Therefore, we only evaluate paths 
with $s$ or $t$ embedded during this subphase.
By design, paths from $s$ previously placed to $t$ placed during 
this subphase have a delta value $\geq \delta_{i}^1/3$. Distance-decreasing paths 
from $s$ placed in this subphase to $t\in G_{i+1}'$ take two 
different directions. If $s$'s neighbor $u$ which is closer to $t$
is on semi-circle $i+1$ then points that are closest to the 
perpendicular bisector to $su$ are along the perimeter of the sector formed
by $s$, $u$, and the origin. The point closest to the perpendicular bisector 
is where the first semi-circle intersects the sector. We translate this 
point down $R_{i+1} + 2$ units along the perpendicular
bisector, and we have an underestimate for the delta value for
any path beginning with a left/right edge. If $s$'s 
neighbor that is closer to $t$ is on the $i$th semi-circle, then
a down edge is followed. To finish, we evaluate down edges $su$
added during the second subphase. The closest vertex to the 
perpendicular bisector to $su$ on the $u$ side is either $u$, or the
vertex placed in the next clockwise position the $i+1$th semi-circle. 
Translating this point $2R_{i+1}$ units away from $su$ along 
the perpendicular bisector gives us the an underestimate for
paths beginning with $su$. 

This completes the proof for the greedy embedding of $G'$. 
 We call the levels where the on-ramps to heavy paths are embedded 
 \emph{super levels}, 
and all other levels are \emph{baby levels}. There are $n-1$ 
baby levels between consecutive super levels and, since any path from root to leaf
in a depth tree travels through $O(\log n)$ different heavy paths, there are $O(\log{n})$ super 
levels. 
\ifFull %
\begin{figure}[!t]
\begin{center}
\begin{tabular}{ccc}
\includegraphics[scale=0.8,viewport=2.0cm 2.0cm 9.5cm 7.5cm,clip=true]{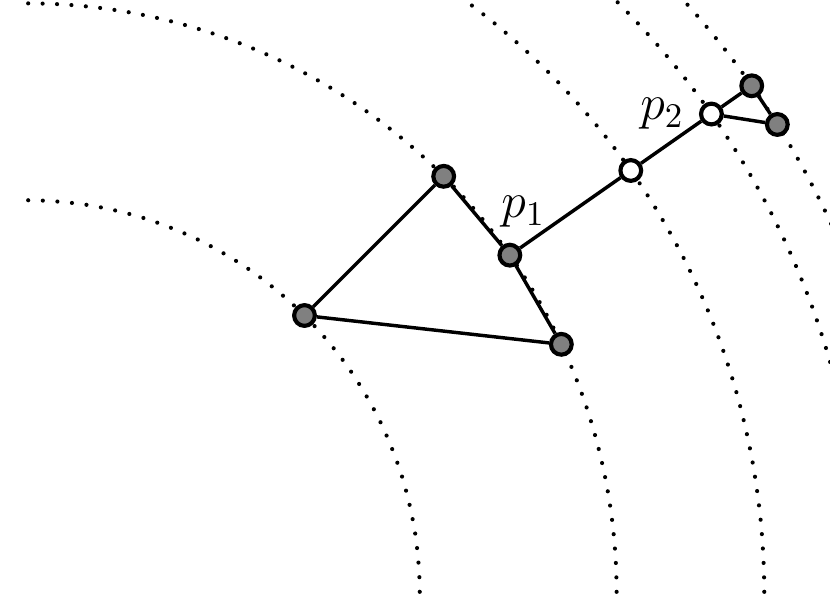}
&
\hspace{10pt}
&
\includegraphics[scale=0.8,viewport=2.0cm 2.0cm 9.5cm 7.5cm,clip=true]{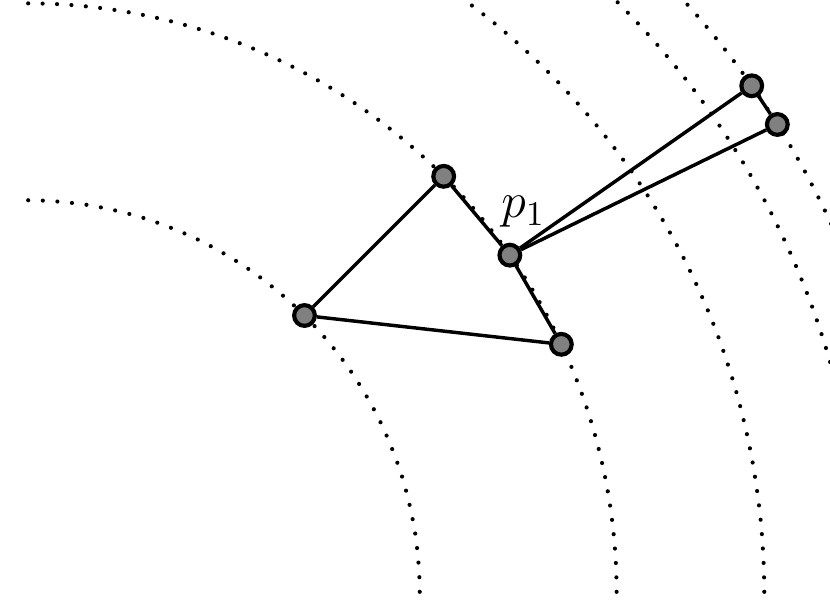} \\
(a) && (b)
\end{tabular}
\caption{(a) Before removal of dummy nodes and (b) after removal.}
\label{fig:removal}
\end{center}
\end{figure}

\subsection{Obtaining a Greedy Embedding of $G$}

Let $G'$ be a modified Christmas cactus graph greedily embedded using the 
procedure discussed above. We now show that collapsing the dummy edges 
leaves us with a graph $G$ and a greedy embedding of $G$. 

Let $C_1$, $C_2$ be any two cycles with a path of dummy edges between 
them. We show that collapsing this path down to a single vertex gives 
us new graph that is also greedily embedded.

\begin{proof}
Assume, without loss of generality, that $C_2$ is a descendant of $C_1$.
Let $P$ be the path of dummy edges between $C_1$ and $C_2$. Let $p_1$ be the vertex 
that cycle $C_1$ shares with $P$, let $p_2$ be the vertex that cycle $C_2$ 
shares with $P$. 

Collapse the path $P$ down to the vertex $p_1$, call this new graph $G''$. 
We assign vertices in $G''$ the same coordinates in $\R^2$ that they are 
assigned in the embedding of $G'$. (See Fig.~\ref{fig:removal}.) 
We show that distance-decreasing paths exist between all pairs of vertices 
in the embedding of $G''$, using the greediness of the embedding of $G'$.

Consider any two vertices $s$ and $t$ in $G''$. There are four cases:

\begin{enumerate}

\item If a distance-decreasing path from $s$ to $t$ in $G'$ involves both 
$p_1$ and $p_2$, then there is a distance-decreasing path in $G''$ 
by transitivity.

\item If a distance-decreasing path from $s$ to $t$ in $G'$ involves $p_1$ 
and not $p_2$, then the same distance-decreasing path exists in $G''$
since no vertices or edges on this path were modified.

\item If a distance-decreasing path from $s$ to $t$ in $G'$ involves $p_2$ and 
not $p_1$, then either $s$ or $t$ is not in $G''$. Therefore, this case 
is irrelevant.

\item If a distance-decreasing path from $s$ to $t$ in $G'$ involves neither 
$p_1$ nor $p_2$, then the same distance-decreasing path exists in $G''$
since no vertices or edges on this path were modified.
\end{enumerate}
Therefore, since there are distance decreasing paths between all $s$ and $t$ in
our embedding of $G'$, there are distance-decreasing paths between all $s$ and $t$
in the embedding of our new graph as well.
\end{proof}

Furthermore, every distance-decreasing path in $G''$ looks like the same
path from $G'$, but with vertices in $P \setminus \{p_1\}$ removed.

We apply the above modification algorithm to $G'$ repeatedly, until all
dummy edges are removed. After removing all of the dummy edges in this
way, we have our original graph $G$ and a greedy embedding of $G$.

\else
To obtain a greedy embedding for $G$, we repeatedly collapse dummy edges
in $G'$ until we get $G$. When we collapse an edge $(p,x_1)$, 
where $p$ is the primary node for the 2-cycle, we collapse 
the edge to vertex $p$ in the embedding. Collapsing in the other
direction may break distance decreasing paths through $p$ and neighbors of $p$ 
embedded on the same semi-circle as $p$. After collapsing all such 
dummy edges, we have a greedy embedding of $G$.

\fi

\subsection{Our Coordinate System}
Let $v$ be a vertex in $G$. We define $\level(v)$ to be the number 
of baby levels between $v$ and the previous super level 
(zero if $v$ is on a super level) and $\cycle(v)$ to be the 
position, $0$ to $2n$, where $v$ is placed when its 
cycle is embedded. These values can be assigned to vertices
without performing the embedding procedure.

Let $s$ be $v$'s ancestor on the first super level. 
The path from $s$ to $v$ passes through $O(\log(n))$ heavy paths, entering
each heavy path at an on-ramp, and leaving at an off-ramp. 
We define $v$'s coordinate to be a $O(\log{n})$-tuple
consisting of the collection of $(\level(\cdot), \cycle(\cdot))$ pairs for each 
off-ramp where a change in heavy paths occurs on the path from $s$ to $v$, and the pair
$(\level(v),\cycle(v))$, which is either an off-ramp or a turnpike. 
Using the coordinate for $v$ and the parameter $n$, we can compute the
Euclidean coordinates for all the turnpikes and off-ramps where a change
in heavy path occurs on the path from $s$ to $v$, including the 
coordinate for $v$. Thus, we have defined a coordinate system for 
the Euclidean plane.

Using a straightforward encoding scheme, each level-cycle pair is encoded 
using $O(\log{n})$ bits. Since a coordinate contains $O(\log{n})$ of 
these pairs, we encode each coordinate using $O(\log^2{n})$ bits. 

\subsection{Greedy Routing with Coordinate Representations}
\ifFull
Although contrived, it is possible to perform greedy geometric
routing by converting our coordinates to Euclidean points and using
the Euclidean $L_2$ metric whenever we need to make a comparison 
along the greedy route. Alternatively, we can define a comparison rule,
which can be used for greedy routing
in our coordinate system, and which evaluates consistently with the 
$L_2$ metric for all vertices on the path from start to goal.
\fi

By design, the routing scheme discussed in Sect.~\ref{section:greedy-routing}
is greedy for our embedding. We develop a comparison rule 
using the potential number of edges that may be traversed on
a specific path from $s$ to $t$.

Let $s_i$ be the vertex between super levels $i$ and $i+1$, whose 
$\level$-$\cycle$ pair is in position $i$ of $s$'s coordinate.
We define $t_i$ similarly. Let $\superlevel(s)$ be the position that 
contains the level-cycle pair for $s$ itself.
Let $h$ be the smallest integer such that $s_h$ and $t_h$ differ.
Using the level-cycle pairs for $s_h$ and $t_h$, we
can compute the level-cycle pair for the off-ramps on the least
common ancestor $C$ that diverge toward $s$ and $t$, which we 
call $s_C$ and $t_C$. That is, if 
$\level(s_h) = \level(t_h)$ then $s_C = s_h$ and
$t_C = t_h$. Otherwise, assume without loss of generality that
$\level(s_h) < \level(t_h)$, then $s_C$'s pair is $(\level(s_h),\cycle(s_h))$
and $t_C$ is a turnpike with the pair $(\level(s_h), n)$. 

We define $l$, $r$, $d$, $u$ be the potential number of left, 
right, down, and up edges that may be traversed from $s$ to $t$. Values 
$d$ and $u$ are simply the number of semi-circles passed through by 
down and up hops, respectively. That is,

{\small
\[d = (\superlevel(s)\cdot n + \level(s)) - (hn + \level(s_C))\]
\[u = (\superlevel(t)\cdot n + \level(t)) - (hn + \level(t_C)).\]
}

If $\cycle(t_C) < \cycle(s_C)$, then we count the maximum number left 
edges on the path from $s$ to $t_C$, and the maximum number of right 
edges from $t_C$ to $t$. That is,

{\small
\[l = \begin{cases}
       \cycle(s) + 2n(d-1) + \cycle(s_C) - \cycle(t_C)&\text{if $s \neq s_C$}, \\
       \cycle(s_C) - \cycle(t_C) &\text{if $s = s_C$}. \end{cases}\]

\[r = \begin{cases}
       2n(u-1) + \cycle(t) &\text{if $t \neq t_C$}, \\
       0 &\text{if $t = t_C$}. \end{cases}\]
}

If $\cycle(t_C) \geq \cycle(s_C)$, then we count the maximum number of right 
edges on the path from $s$ to $t_C$, and the maximum number of right edges 
from $t_C$ to $t$. That is,

{\small
\[l = 0\]
\[r = r_1 + r_2,\text{ where}\]
\[r_1 = \begin{cases}
       2n-\cycle(s) + 2n(d-1) + \cycle(t_C) - \cycle(s_C) &\text{if $s \neq s_C$}, \\
       \cycle(t_C) - \cycle(s_C) &\text{if $s = s_C$}. \end{cases}\]
\[r_2 = \begin{cases}
        2n(u-1) + \cycle(t) &\text{if $t \neq t_C$}, \\
        0&\text{if $t = t_C$}. \end{cases}\]
}

Our comparison rule is:
\[D(s,t) = l + r + (2n+1)u + d.\]

Following the routing scheme from Sect.~\ref{section:greedy-routing}, any 
move we make toward the goal will decrease $D(\cdot,\cdot)$, and all other
moves will will increase $D(\cdot,\cdot)$ or leave it unchanged. 
Therefore, we can use this comparison rule to perform greedy routing 
in our embedding efficiently, and comparisons made along the greedy route
will evaluate consistently with the corresponding Euclidean coordinates 
under the $L_2$ metric.

\section{An Optimal Succinct Greedy Embedding}
Conceptually, the $\level(\cdot)$ and $\cycle(\cdot)$ values used
in the previous section are 
encoded as integers whose binary representation corresponds to
a path from root to a leaf in a full binary tree with $n$ leaves. Instead 
of encoding with a static $O(\log{n})$ bits per integer, we will 
modify our embedding procedure so we can further exploit the heavy path 
decomposition of the dual tree $T$, using 
\emph{weight-balanced binary trees}~\ifFull\cite{GilMoo-BSTJ-59,Knu-AI-71}.\else%
\cite{Knu-AI-71}.\fi

\begin{definition} A {\bfseries weight-balanced binary tree} is a binary tree
which stores weighted items from a total order in its leaves. If item $i$
has weight $w_i$, and all items have a combined weight of $W$ then item 
$i$ is stored at depth $O(\log{W/w_i})$. An inorder listing of the leaves outputs
the items in order.
\end{definition}

\ifFull By using appropriate weight functions with our weight-balanced 
binary trees, we will be able to get telescoping sums
for the lengths of the codes for the $\level(\cdot)$ 
and $\cycle(\cdot)$ values, 
giving us $O(\log{n})$ bits per coordinate, which is optimal. \fi

\subsection{Encoding the Level Values}
\ifFull
As in the $O(\log^2{n})$ embedding, we will lay the heavy paths between 
super levels. However, we no longer require the on-ramps of heavy paths to be
embedded on super levels, nor do we require adjacent cycles on the same 
heavy path to be embedded on consecutive levels; instead, cycles will be
assigned to baby levels by an encoding derived from a 
weight-balanced binary tree. \fi

\ifFull
We will have a different weight-balanced binary tree for each heavy path
in our depth tree. 
The items that we store in the tree are the cycles on the heavy path. 
The path in the weight-balanced binary tree from the root to 
the leaf containing a cycle gives us an encoding for the $\level$ that the 
cycle should be embedded on between super levels. 
\fi

Suppose we have a depth tree $T$ for $G$, and a heavy path 
decomposition of $T$. Let $C$ be a simple cycle in $G$ on some heavy path
$H$ and let $C_{\mathrm{next}}$ be the next cycle on the heavy path $H$,
if it exists. Let $n(C)$ be the number of vertex descendants of 
$C$ in $G$. We define a weight function $\gamma(\cdot)$ on the cycles 
in $G$ as follows:
{\small
\[
\gamma(C) = \begin{cases}
         n(C)&\text{if $C = \tail(H)$}, \\
         n(C) - n(C_{\mathrm{next}})&\text{if $C \neq \tail(H)$}.
         \end{cases}
\]
}
\ifFull \noindent That is, $\gamma(C)$ is the number of descendants of cycle $C$ in $G$ 
excluding the descendants of the next cycle on the heavy path with $C$. \fi

For each heavy path $H$, create a weight-balanced binary tree $B_H$ containing
each cycle $C$ in $H$ as an item with weight $\gamma(C)$, and impose a total order 
so that cycles are in their path order from $\head(H)$ to $\tail(H)$. 

Let $v$ be a vertex whose coordinate we wish to encode, and suppose $v$ is located
between super levels $l$ and $l+1$. Let $v_i$ be the vertex whose $\level$-$\cycle$
pair is in position $i$ of $v$'s coordinate. Let $v_i$ be contained in cycle $C_i$ 
(such that $v_i$ is not $C_i$'s primary node) on heavy path $H_i$. 
\ifFull
Then the coordinate for $v$ will contain the collection of 
$\level(\cdot)$ values for each 
off-ramp $v_i$ on the path to $v$, and the $\level(\cdot)$ value for $v$
itself. Let $C_i$ be the cycle containing vertex $v_i$, such 
that $v_i$ is not the primary node for $C_i\in H_i$. \fi 
The code for $\level(v_i)$ is a bit-string representing the path 
from root to the leaf for $C_i$ in the weight-balanced binary tree 
$B_{H_i}$. Let $W_i$ be the combined weight of the items in $B_{H_i}$. 
Since $C_i$ is at a depth of $O(\log{W_i/\gamma(C_i)})$, this is 
length of the code.  
Thus, the level values in $v$'s coordinate are encoded with
$O(\sum_{0\leq i\leq l}\log{W_i/\gamma(C_i)})$ bits total. \ifFull
We now show that this is a telescoping sum, giving us $O(\log{n})$ bits total.
\ifFull
All descendants counted in $W_i$ are counted in $\gamma(C_{i-1})$, 
therefore, we have that $\gamma(C_{i-1}) \geq W_i$. By subtracting off  
descendants that are further along the heavy path, we ensure 
that $W_0 = n$. Thus, $\sum_{0\leq i\leq l} \log{W_i/\gamma(C_{i})} 
\leq \log{W_0/\gamma(C_l)} \leq \log{n}$. \else
By design, this sum telescopes to $O(\log{n})$ bits.
\fi

\subsection{Encoding the Cycle Values}
For a node $v$ in $G$ we define a weight function $\mu(v)$ to be the number
of descendants of $v$ in $G$.

Let $C=(p,x_1,x_2,\ldots,x_m)$ be a cycle in $G$, where $p$ is 
the primary node of $C$. Let $x_h$ be the turnpike that connects 
$C$ to the next cycle on the heavy path, if it exists. 
Let $x_i$ have weight $\mu(x_i)$ and impose a total order so 
$x_j<x_k$ if $j<k$. For each cycle $C$, we create a weight-balanced binary tree $B_C$ 
containing nodes $x_1$ to $x_m$ as follows. We first create 
two weight-balanced binary trees $B^1_C$ and $B^2_C$ 
where $B^1_C$ contains $x_j$ for $j < h$ and $B^2_C$ contains
 $x_k$ for $k > h$. If no such $x_h$ exists, then choose an 
integer $1\leq k\leq m$ and insert items $x_j$ for $j<k$ into 
$B^1_C$ and insert the remaining items into $B^2_C$. We form our single 
weight-balanced binary tree $B_C$ in two steps: (1) create a tree $B^3_C$ with 
$B^1_C$ as a left subtree and a node for $x_h$ as a right subtree, and (2)
form $B_C$ with $B^3_C$ as a left subtree and $B^2_C$ as a right subtree.
We build $B_C$ in this way to ensure that every turnpike is given the same
path within its tree, and hence the same cycle code and value.

The code for $\cycle(v_i)$ is a bit-string representing the path from root to the 
leaf for $v_i$ in the weight-balanced binary tree $B_{C_i}$. Let $W_i$ be the
combined weight of the items in $B_{C_i}$. Since $v_i$
is at a depth of $O(\log{W_i/\mu(v_i)})$, this is length of the code.  
Thus, the $\cycle$ values in $v$'s coordinate are encoded
with $O(\sum_{0\leq i \leq l}\log{W_i/\mu(v_i)})$ bits total. 
\ifFull We now show that this is a telescoping sum, giving us $O(\log{n})$ 
bits total.

Every descendant counted in $W_i$ is also counted in $\mu(r_{i-1})$, 
thus $\mu(r_{i-1}) \geq W_i$. By design, $W_0 = n$. Hence 
$\sum_{0\leq i\leq l} \log{W_i/\mu(r_{i})} \leq \log{W_0/w(r_l)} \leq \log{n}$.
\else
By design, this sum telescopes to $O(\log{n})$ bits.
\fi

\subsection{Interpreting the Codes}

Let $c$ be the smallest integer constant such that item $i$ stored 
in the weight-balanced binary
tree is at depth $\leq c\log{W/w_i}$. We can treat the position of $i$ 
in the weight-balanced binary tree as a position in a full binary tree of 
height $c\log{n}$. We interpret this code to be the number of tree nodes 
preceding $i$ in an in-order traversal of the full binary tree. Using our
codes as described, we require $2n^c - 2$ baby levels between each super 
level and $8n^c - 1$ cycle positions.

\subsection{An Overview of the Optimal Embedding}
Let $T$ be the depth tree for our Christmas cactus graph $G$. 
We create weight-balanced binary trees on the heavy paths in $T$ and
on each of the cycles in $G$, giving us the $\level$ and $\cycle$ codes for
every vertex. We adjust the graph modification procedure 
so that adjacent cycles on heavy paths are spaced out according to the
level codes. That is, adjacent cycles on the same heavy path have
heavy dummy edges (dummy edges that are considered to be on 
the heavy path) inserted between them so that they are placed on 
the appropriate baby levels. For cycles on different heavy paths, 
we insert dummy edges to pad out to the next superlevel, 
and heavy dummy edges to pad out to the appropriate baby level. 

We embed the modified graph analogously to our $O(\log^2n)$ embedding, except
that the cycle codes dictate vertex placements.
We augment our coordinate system to store the $\level$ value 
for elements on the root cycle, otherwise it is not possible to compute the 
corresponding Euclidean point from our succinct representation.
The same comparison rule applies to our new coordinate system, with little
change to account for the new range of $\level$ and $\cycle$ values. Using this 
embedding scheme and coordinate system we achieve optimal 
$O(\log n)$ bits per coordinate.

\ifFull
\section{Conclusion}
We have provided a succinct coordinate-based representation for the
vertices in 3-connected planar graphs so as to support greedy routing in
$\R^2$.
Our method uses $O(\log n)$ bits per vertex and allows greedy routing 
to proceed using only our representation, in a way that is consistent
with the Euclidean metric.
For future work, it would be interesting to design an efficient
distributed algorithm to perform such
embeddings.
\fi

\bibliographystyle{abbrv}
\bibliography{geom,greedy,roads}

\end{document}